# Ultrafast dynamics of wavelength-sensitive magnons in unconventional compensated semiconducting antiferromagnet

Hanshen Huang†[1], Tao Qu†[1], Yang Cheng[1], Lixuan Tai[1], Christopher Eckberg[1], Quanjun Pan[1], Abdullah Alrasheed[1], Su Kong Chong[1], Bingqian Dai[1], Yaochen Li[1], Qingyuan Shu[1], Chao-Yao Yang[2], Jie-Xiang Yu[3], Gen Yin[4], and Kang L. Wang*[1]

[1]Department of Electrical and Computer Engineering, Department of Physics and Astronomy, and Department of Material Science and Engineering, University of California, Los Angeles, California 90095, USA
[2]Department of Material Science and Engineering, National Yang Ming Chiao Tong University, Hsinchu 300044, Taiwan
[3]School of Physical Science and Technology, Soochow University, Suzhou 215006, China
[4]Department of Physics, Georgetown University, Washington, D.C. 20057, USA

**Abstract:**

Antiferromagnet is a promising candidate for the next generation spintronic devices, benefiting from its ultrafast dynamics and spontaneous zero stray field. However, the understanding of their ultrafast spin behaviors is lacking due to the challenges of controlling/detecting the quenched net magnetization. Unconventional compensated semiconducting antiferromagnets present strong time-reversal symmetry breaking, spin splitting in the momentum space, and suitable bandgap for optical control/detection. Thus, it is a powerful platform to uncover the ultrafast dynamics of antiferromagnets. Here, we show an exotic wavelength-dependent spin dynamic in the unconventional compensated semiconducting antiferromagnet $\alpha$-MnTe via time-resolved quadratic magneto-optical Kerr effect measurement, where the probing photon energy of the laser matches its bandgap. This direct excitation and detection of distinct magnon modes reveal varying spin behaviors and time characteristics in a broad temperature range. It originates from the spins triggered at different bands of electronic structures and is depicted in an energy transfer model among electrons, phonons, and magnons. Our study of exotic optical properties in this unconventional semiconducting antiferromagnet fulfills the missing information of spin evolution in the time domain and paves the way for its utilization in ultrafast spintronic devices.

**Introduction:**

Spintronic device, utilizing the electron's spin degree of freedom, is a promising candidate in the new paradigm to push CMOS (complementary metal-oxide-semiconductor) to the quantum limit. Antiferromagnet, a material base of spintronic devices, consists of sets of two pointing spins quenching each other where the coupled spin sets push the operation frequency to terahertz[1,2]. Additionally, the compensated spin sets cause the absence of the stray field and could result in a higher areal density of devices compared with ferromagnet/ferrimagnet-based spintronic devices. These features give antiferromagnets great advantages in ultrafast spintronic devices, such as magnetic random-access memory, that may scale with modern CMOS technology[3,4]. However, the compensated magnetic moment in antiferromagnets resulting in a zero net magnetic moment, posts a challenge in probing or controlling the spin state, and obscures the fundamental understanding of the spin dynamics in an ultrafast time scale, resulting in a lag in the development of antiferromagnet-based spintronic devices.

Recent research has adopted electrical and optical approaches to investigate the spin behavior in antiferromagnets. Electrical control and detection [5,6] of Néel vector in antiferromagnets is in debate. For example, 8-leg Hall bar device measurement has indirect access to spin information, and thus the obtained signal is often made complicate by inevitable Joule heating[7]. Optical imaging antiferromagnetic domains using magnetic linear birefringence effect[8,9], directly shows the spatial configuration, but does not to provide the spin evolution in the time domain. Using the time-resolved optical pump-probe technique can resolve the spin dynamics in sub-picosecond or a much finer time scale. This time-resolve technique has mainly focused on metallic antiferromagnets such as CuMnAs[10] and $Mn_3Sn$[11], where the photon energy of the laser is not a concern due to the metallic continuous band structure. For insulators with a wide bandgap[12–14] such as NiO (bandgap ~3.6 eV), indirect optical manipulation of the antiferromagnetic order at ultrafast time scale was achieved by the interfacial energy transfer from the capped non-magnetic metal layer to the antiferromagnetic insulator. Since direct manipulation requires the photon energy to match the bandgap, it is a challenge to employ the time-resolved technique in an insulator due to the need of short laser wavelength[9]. In addition, many exotic properties of noncollinear antiferromagnets[3,4] originate from the broken symmetry of the spin degeneracy and thus the momentum-dependent spin polarization. These studies could lay the foundation for the development of ultrafast spintronic devices but has not been demonstrated to dilucidated their ultrafast features.

$\alpha$-MnTe (space group P6$_3$/mmc), the material we will focus in our study, is an unconventional compensated semiconducting antiferromagnet[15]. Although it is a collinear antiparallel magnetic ordering, its unconventional phase introduces strong time-reversal symmetry breaking and alternating spin polarization in the real-space crystal structure and the momentum-space electronic structure. This lifting of Kramers spin degeneracy also predicts its sensitivity to optical approaches [16,17],,e.g. magneto-optical coupling, but hasn't been studied yet. In this work, we demonstrate ultrafast spin dynamics in this unconventional antiferromagnet through a time-resolved quadratic magneto-optical Kerr effect (TR-QMOKE) technique, where the photon energy of the laser matches the bandgap of this semiconductor. Thus, our direct excitation and detection of spins by the laser reveal distinct spin behaviors under different photon energies, indicating the ultrafast dynamics of magnons distributed across the valence band with unseen interesting temporal features. These features persist in a wide temperature range, even above the Néel temperature when the weak antiferromagnetic order exists. Our work explores this promising unconventional antiferromagnet, characterizes the time scales of the ultrafast dynamics through the direct optical method, and addresses the missing part of the foundation for ultrafast antiferromagnetic spintronic devices.

**Results:**

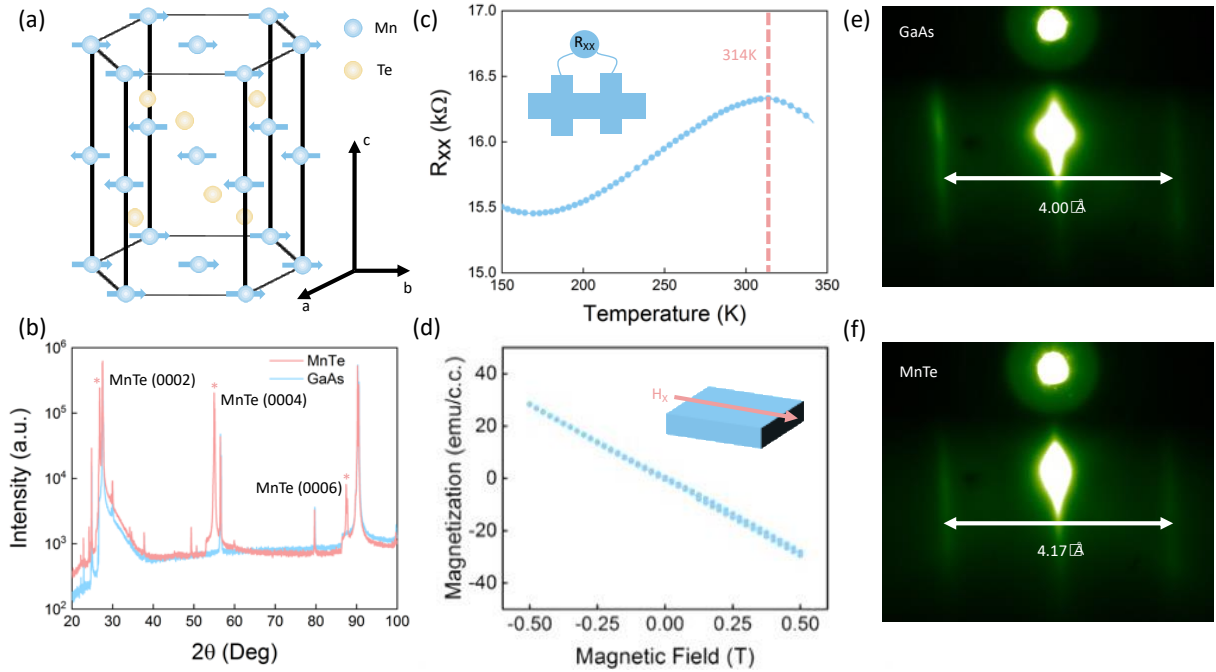

**Figure 1. Material characterization. (a)** Crystallographic and magnetic structure of $\alpha$-MnTe (P6$_3$/mmc), where blue balls stand for Mn atoms and yellow balls stand for Te atoms. **(b)** XRD signal for GaAs substrate and $\alpha$-MnTe showing good crystallinity of $\alpha$-MnTe thin film. The small side peaks are from the impurity of the X-ray source. **(c)** Temperature-dependent measurement of longitudinal resistance showing a peak at Néel temperature. **(d)** SQUID measurement at room temperature with an in-plane field. **(e)** and **(f)** RHEED pattern before and after $\alpha$-MnTe growth indicating 2-D single crystalline growth of the thin film.

The particular phase used in the study is $\alpha$-MnTe, which has a hexagonal close-packed structure with antiferromagnetic ordering along the c-axis (**Fig. 1a**). The $\alpha$-MnTe thin film grown by molecule beam epitaxy (MBE) on GaAs(111) substrate is 35 nm thick with a 2 nm (Bi,Sb)$_2$Te$_3$ buffer layer to ensure the formation of hexagonal NiAs phase of MnTe ($\alpha$-MnTe). The quality and thickness of $\alpha$-MnTe are monitored by *in situ* reflection high-energy electron diffraction (RHEED) while growing the sample. The streaky patterns shown in **Fig. 1e** and **Fig. 1f** before and after growth indicate the 2-D single-crystalline growth of the thin film. Material characterization then is performed using X-ray and transport to verify the quality of our MBE-grown $\alpha$-MnTe thin film. As is shown in **Fig. 1b**, the X-ray diffraction (XRD) pattern shows only (00X) (X=2, 4, 6) peaks of α-MnTe and peaks from GaAs substrate, which confirms the good crystallinity of the thin film. Transport measurement is performed to determine the Néel temperature of the MnTe thin film. As is shown in **Fig. 1c**, the peak point of the temperature-dependent transverse resistance curve indicates the transition temperature (Néel temperature) at 314 K which is similar to the results from literature[18]. Superconducting quantum interference device (SQUID) measurement is conducted with an in-plane field under room temperature to verify the antiferromagnetic origin of MnTe. The near-zero magnetic moment at the zero field and disappearance of the hysteresis loop during field sweeping further confirm the antiferromagnetic nature of our MBE-grown MnTe thin film (**Fig. 1d**).

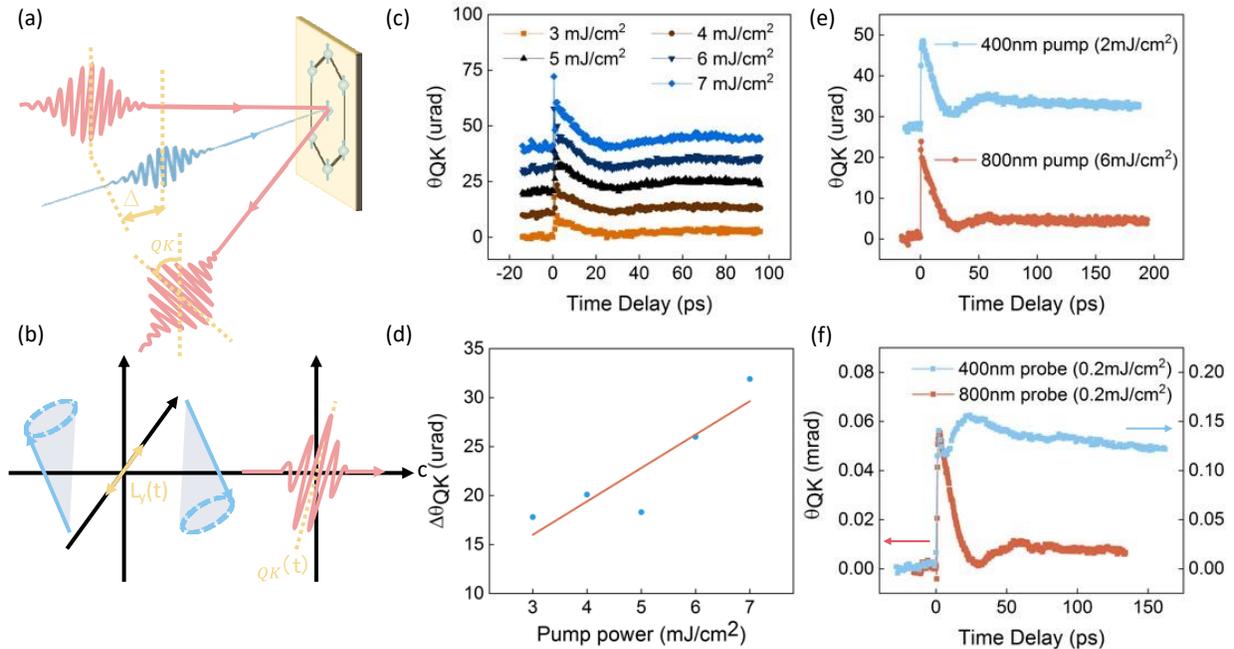

**Figure 2. TR-QMOKE measurement. (a)** Schematics of pump-probe measurement. **(b)** Schematics of mechanism for detecting spin dynamics using QMOKE. **(c)** and **(d)** Pump power-dependent pump-probe measurement with an 800nm pump and an 800nm probe. **(e)** Pump wavelength-dependent measurement with the 800nm/400nm pump light and the 800nm probe light. **(f)** Probe wavelength-dependent measurement with the 800nm/400nm probe light and the 800nm pump light.

TR-QMOKE measurement reaping the power of both high order magneto-optical Kerr effect that detects the antiferromagnetic order and the time-resolved pump-probe technique that detects the evolution of the order. **Fig. 2a** shows the schematics of the TR-QMOKE measurement. As is shown in this figure, a pump beam (blue) is used to excite the $\alpha$-MnTe thin film, and a probe beam (red) is used to probe the Néel order changes due to the excitation. The probe beam is normally incident to the (0001) plane of the $\alpha$-MnTe thin film where the Néel vector lies in the sample plane. When the pump laser pulse is first applied at time zero, the transient heat generated heats the $\alpha$-MnTe far above the Néel temperature and causes transient changes of the antiferromagnetic order. After the excited laser pulse passes through the sample, the temperature of the sample drops and gradually evolves back to its equilibrium state. This change of the Néel vector is shown in the change of linear polarization of the reflected probe light (**Fig. 2b**). The signal of the TR-QMOKE signal is shown in **Fig. 2c** for different levels of the pump power. The figure shows that as the pump power increases from 3mJ/cm² to 7mJ/cm², the signal also increases linearly with the pump power (**Fig. 2d**). Consequently, the TR-QMOKE signal can be expressed by the following equation[10]:

$$\theta_{QK} = \beta sin2\varphi \delta M(\Delta t), \quad (1)$$

where $\beta$ is the magneto-optical coefficient, $\varphi$ is the angle between the light polarization and the Néel vector, and $\delta M$ is the pump-induced magnetization change at time $\Delta t$. A wavelength-dependent measurement of both the pump and the probe beams is performed to see how photons with different energies interact with $\alpha$-MnTe thin film. The chosen photon energies are 1.55 eV (800 nm) and 3.1 eV (400 nm). This is because 1.55 eV matches closely with the bandgap of $\alpha$-MnTe and 3.1 eV is above the bandgap across the whole k-space. As is shown in **Fig. 2e**, the probe beam remains unchanged at the same 800nm while the pump light is switched between 800nm (6mJ/cm²) and 400nm (2mJ/cm²). The signal shows no observable feature difference under different pump wavelengths, which implies the excitation mechanism is the same no matter what the pump photon energy is, with only difference being the pumping efficiency. However, when the wavelength of the probe beam is switched from 800nm to 400nm, the TR-QMOKE dynamics in $\alpha$-MnTe show different features (**Fig. 2f**). As is shown in the figure, when the wavelength of the probe beam is set at 800nm, the sample returns to equilibrium very fast, which is a behavior similar to other insulating antiferromagnetic materials such as NiO. However, when the pump beam remains at 800nm but the probe beam is switched to 400nm, MnTe returns to the equilibrium much slower and shows a different TR-QMOKE dynamic that has never been reported before, to our knowledge.

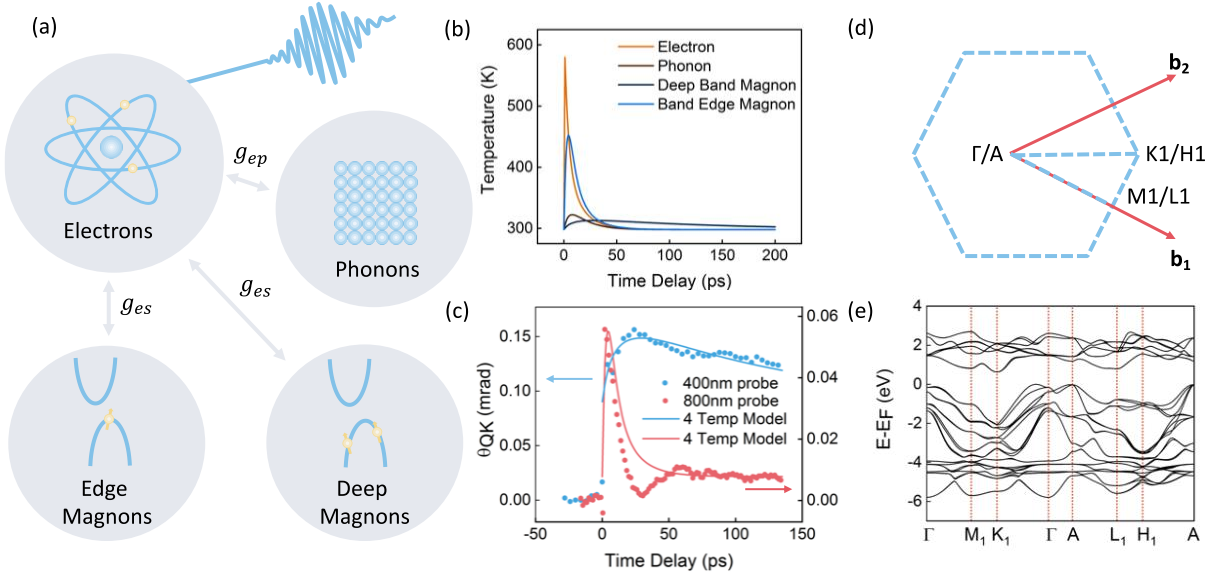

**Figure 3. Four-temperature model**. **(a)** Illustration of the phenomenological temperature model incorporating systems of four quasiparticles: electrons, phonons, band edge magnons and deep band magnons, and their interactions. **(b)** Four-temperature model showing temperature evolution with time for the different sub-systems respectively. **(c)** The magnon temperatures from the four-temperature model: band edge and deep band magnons, fitting with the experimental TR-QMOKE signal. **(d)** The path of the high symmetry points in the k-space where the band energy is calculated. **(e)** $\alpha$-MnTe band diagram via an *ab initio* calculation.

To better understand and explain the origin of this effect, we develop a four-temperature model (illustrated in **Fig. 3a**) listed below, based on the well-known three-temperature model which phenomenologically describes the energy transfer among laser excitation, electrons, magnons, and phonons[19]

$$\eta \frac{dT_e}{dt} = g_{ep}(T_p - T_e) + g_{es}(T_{s1} - T_e) + g_{es}(T_{s2} - T_e) + P(t), \quad (2)$$

$$C_p \frac{dT_p}{dt} = g_{ep}(T_e - T_p) - \kappa(T_p - T_{amb}), \quad (3)$$

$$C_{s1} \frac{dT_{s1}}{dt} = g_{es}(T_e - T_{s1}), \quad (4)$$

$$C_{s2} \frac{dT_{s2}}{dt} = g_{es}(T_e - T_{s2}), \quad (5)$$

where $\eta$ is the heat capacity of electron, $T_e$ is the electron temperature; $g_{ep}$ is the electron-phonon coupling coefficient; $T_p$ is the phonon temperature; $g_{es}$ is the electron-magnon coupling coefficient; $T_{s1}$ ($T_{s2}$) is the temperature for magnons at the valence band edge (deep in the valence band); P(t) is the energy source term accounting for the laser excitation; $C_p$ is heat capacity of the phonons; $\kappa$ is the thermal boundary conductance; $T_{amb}$ = 300 K is the constant temperature of the heat sink; $C_{s1}$ ($C_{s2}$) is the heat capacity of magnons in the band edge (deeper band) branch. The three coupling constants $g_{ep}$, $g_{es}$, and $\kappa$ are taken independent of temperature, where the values of these constants are adopted from literature reports[14]. A more detailed explanation of the four-temperature model can be found in supplementary note S1. The time evolution of four subsystems' temperature is shown in the **Fig. 3b**. The model clearly reveals a fast decade band edge magnon mode and a slow decade deep magnon mode. And the model fits well with the experimental data shown in **Fig. 3c**.

To better support our four-temperature model, $\alpha$-MnTe band diagram following the path of high symmetry points in the k-space (**Fig. 3d**) is obtained from the *ab initio* calculation[20] with the lattice constant obtained from the RHEED pattern (**Fig. 3e**). As described previously, the pump beam excitation drives the sample into a non-equilibrium state and the probe beam detects its relaxation back to the equilibrium as a function of time. With the pump beam wavelength set at 400 nm, the corresponding photon energy (3.10eV) is much larger than all the direct bandgap of $\alpha$-MnTe. While for the pump wavelength of 800nm, the corresponding photon energy (1.55eV) is only comparable to the direct bandgap around Γ(A) point. Thus, the pump process is wavelength independent but with a higher efficiency at 400nm, which is consistent with experiment observations. Following the pump beam exciting the magnons, the probe beam detects the dynamics of magnons in the time domain. The probe signal is based on QMOKE, proportional to the magnon numbers. And the QMOKE originates from the difference of the complex refractive indices which is determined by the electronic band structure. A detailed explanation can be found in supplementary note S2. Consequently, for the probe wavelength of 800nm where its photon energy is 1.55 eV, it only detects the magnons that couples to the electrons at the valence band edge near Γ(A) point. On the other hand, the probe wavelength of 400 nm with corresponding photon energy 3.10 eV, overcomes the bandgap over all the k-space and can detect the magnons that couples to electrons both at the valence band edge and deep in the valence band, especially near the $K_1$ resonance point. Consistent with the ultrafast experiment, the probe wavelength of 400 nm gives richer features in the dynamics, which is an overlay of the time-evolving of two magnon branches. Through the resonance between the probe photon and the magnon coupling to electrons deep in the valence band, the fine change of the magnon deep in the valence band is more sensitively detected.

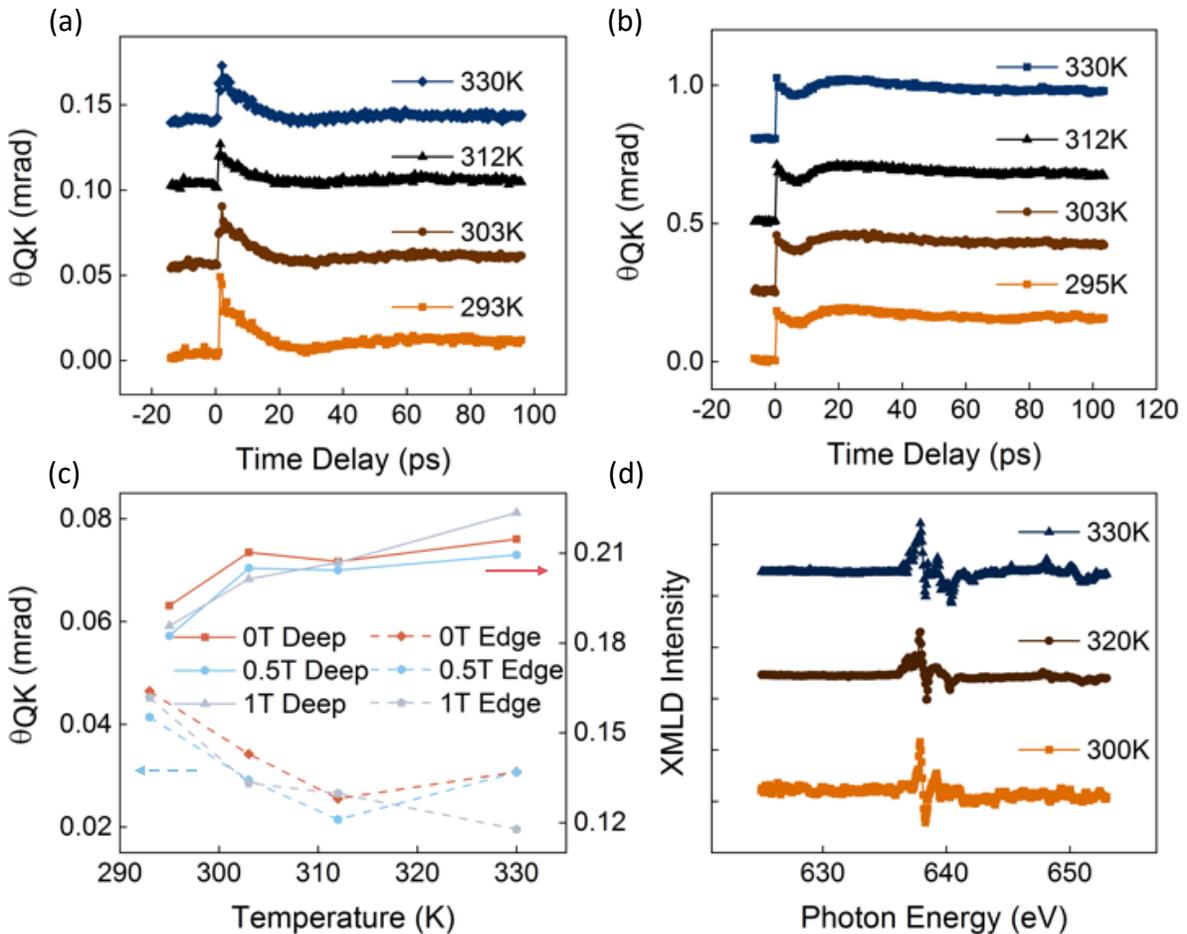

**Figure 4. Temperature-dependent behavior of the edge and deep magnons. (a)** TR-QMOKE measurements under the 800nm probe at different temperatures (293K, 303K, 312K, and 330K) without an external magnetic field. **(b)** TR-QMOKE measurements under the 400nm probe at different temperatures (295K, 303K, 312K, and 330K) without an external magnetic field. **(c)** Temperature-dependent behavior with 400nm (800nm) probe under different in-plane magnetic fields. **(d)** Temperature-dependent XMLD results.

Temperature and field-dependent measurements are then conducted under 10mJ/cm$^2$ pump power with an external magnetic field in the sample plane. **Fig. 4a** and **Fig. 4b** show the temperature-dependent measurement probing by the 800nm and 400nm beams, respectively. The $\theta_{QK}$ peak height is extracted and plotted in **Fig. 4c**, showing that as the temperature increases, the $\theta_{QK}$ peak decreases for the 800nm probe while the $\theta_{QK}$ peak increases when probed at 400nm. The data do not change significantly with the external in-plane magnetic field up to 1 T. One thing to notice is that there is a noticeable $\theta_{QK}$ signal when the temperature is larger than the Néel temperature up to 330K. This is because there is a weak Néel order above the Néel temperature. This is proved by X-ray magnetic linear dichroism (XMLD) experiments, showing the signal up to 330K, as illustrated in **Fig. 4d**.

**Discussion:**

Unconventional compensated semiconducting antiferromagnet $\alpha$-MnTe is a promising candidate for future antiferromagnetic applications such as random-access memory. It shows an antiferromagnetic order up to 330K and is strongly immune to any external magnetic field for 1T. The antiferromagnetic spin dynamics within $\alpha$-MnTe can be effectively read through the TR-QMOKE method and has an exotic wavelength dependence behavior. The optical probe can detect the spins at different parts of the electron band structure when photon energy matches the optical transition in the semiconductor band structure. The different spin dynamics under different probe beam wavelengths can be explained by the fact that two types of magnon coexist under transient thermal excitation and picked up by different photon energies respectively. Although further research is required to obtain a more detailed microscopic picture of $\alpha$-MnTe's spin system, this work provides insight for future spintronic applications using this unconventional antiferromagnet.

**Method:**

Material Growth

The α-MnTe thin films in this paper were grown on epi-ready semi-insulating GaAs (111)B substrates in an ultra-high vacuum, Perkin-Elmer molecular beam epitaxy (MBE) system, with a 2 nm $(Bi,Sb)_2Te_3$ buffer layer in between to ensure the formation of α-phase MnTe instead of other phases. Before growth, the substrates were loaded into the MBE chamber and pre-annealed at the temperature of 630 °C in a Te-rich environment to remove the native oxide on the surface. During the growth of the (Bi,Sb)2Te3 buffer layer, high-purity Bi, Sb, and Te were evaporated from standard Knudsen cells, with the substrate kept at 200 °C. For the growth of the α-MnTe, high-purity Mn and Te were evaporated from standard Knudsen cells. The substrate temperature ramped up from 200 °C to 300 °C in 10 minutes during the growth and then was kept constant at 300 °C. The RHEED *in situ* was used to monitor the growth and thickness of the materials, and the streaky patterns indicate the 2-D single-crystalline growth of the thin film.

TR-QMOKE

Femtosecond pulses (FWHM = 200 fs, 800nm) are generated by Spitfire (Spectra Physics) with a repetition rate of 1KHz. The pulse is then separated into the pump pulse and the probe pulse by a beam splitter. The pump pulse will go through a chopper (37Hz) and incident on the sample surface obliquely. The probe pulse then goes through a delay line and incident on the sample surface normally. The reflected light is separated into the P and S waves by a Wollaston Prism and collected by a Balanced Photodiode. The collected signal is transferred into a Lock-in Amplifier (Stanford Research 380). By changing the time delay between the arrival time of the pump pulse and probe pulse, the time dependent spin dynamic information is extracted.

Wavelength-dependent measurement

Wavelength-dependent measurements are conducted with a frequency doubler by applying a Beta Barium Borate (BBO) crystal. The BBO Crystal is put in the path of the probe beam to convert the 800nm light to 400nm. The 400nm light is then used as the probe beam for the TR-QMOKE measurements.

Temperature-dependent measurement

Temperature-dependent measurements are conducted by applying a positive temperature coefficient (PTC) material. By increasing the input current of the PTC material, the temperature also increases. A thermocouple multimeter is used to monitor the sample surface temperature in real-time. The TR-QMOKE measurement is conducted after the temperature becomes stable after 20 min.

XMLD measurement

X-ray absorption spectra (XAS) at Mn L 2,3 edge were acquired using a linearly polarized X-ray without and with a magnetic field (1 Tesla), denoted as and, respectively, at various temperatures. The X-ray incidence was in a gracing mode to the film surface with 80 degrees to the normal direction. The linearly polarization of the X-ray is set in the film plane and orthogonal to the external magnetic field to ensure gaining the XAS difference between and, based on the spin-flop geometry. The XAS measurements were performed at Beamline 45A2, Taiwan Photon Source (TPS), National Synchrotron Radiation Research Center (NSRRC), Taiwan.

**Reference**

1. Jungwirth, T., Marti, X., Wadley, P. & Wunderlich, J. Antiferromagnetic spintronics. *Nature Nanotechnology 2016 11:3* **11**, 231–241 (2016).
2. Baltz, V. *et al.* Antiferromagnetic spintronics. *Rev Mod Phys* **90**, 015005 (2018).
3. Qin, P. *et al.* Room-temperature magnetoresistance in an all-antiferromagnetic tunnel junction. *Nature 2023 613:7944* **613**, 485–489 (2023).
4. Chen, X. *et al.* Octupole-driven magnetoresistance in an antiferromagnetic tunnel junction. *Nature 2023 613:7944* **613**, 490–495 (2023).
5. Wadley, P. *et al.* Spintronics: Electrical switching of an antiferromagnet. *Science (1979)* **351**, 587–590 (2016).
6. Cheng, Y., Yu, S., Zhu, M., Hwang, J. & Yang, F. Electrical Switching of Tristate Antiferromagnetic Néel Order in α-Fe2 O3 Epitaxial Films. *Phys Rev Lett* **124**, 027202 (2020).

7.  Chiang, C. C., Huang, S. Y., Qu, D., Wu, P. H. & Chien, C. L. Absence of evidence of electrical switching of the antiferromagnetic néel vector. *Phys Rev Lett* **123**, 227203 (2019).

8.  Xu, J. *et al.* Imaging antiferromagnetic domains in nickel oxide thin films by optical birefringence effect. *Phys Rev B* **100**, 134413 (2019).

9.  Xu, J. *et al.* Optical imaging of antiferromagnetic domains in ultrathin CoO(001) films. *New J Phys* **22**, 083033 (2020).

10. Saidl, V. *et al.* Optical determination of the Néel vector in a CuMnAs thin-film antiferromagnet. *Nature Photonics 2016 11:2* **11**, 91–96 (2017).

11. Zhao, H. C. *et al.* Large ultrafast-modulated Voigt effect in noncollinear antiferromagnet Mn3Sn. *Nature Communications 2021 12:1* **12**, 1–8 (2021).

12. Hortensius, J. R. *et al.* Coherent spin-wave transport in an antiferromagnet. *Nature Physics 2021 17:9* **17**, 1001–1006 (2021).

13. Grishunin, K., Mashkovich, E. A., Kimel, A. V., Balbashov, A. M. & Zvezdin, A. K. Excitation and detection of terahertz coherent spin waves in antiferromagnetic. *Phys Rev B* **104**, 024419 (2021).

14. Wust, S. *et al.* Indirect optical manipulation of the antiferromagnetic order of insulating NiO by ultrafast interfacial energy transfer. (2022) doi:10.48550/arxiv.2205.02686.

15. Gonzalez Betancourt, R. D. *et al.* Spontaneous Anomalous Hall Effect Arising from an Unconventional Compensated Magnetic Phase in a Semiconductor. *Phys Rev Lett* **130**, (2023).

16. Šmejkal, L., Sinova, J. & Jungwirth, T. Emerging Research Landscape of Altermagnetism. *Phys Rev X* **12**, (2022).

17. Mazin, I. I. Altermagnetism in MnTe: Origin, predicted manifestations, and routes to detwinning. *Phys Rev B* **107**, 100418 (2023).

18. Kriegner, D. *et al.* Multiple-stable anisotropic magnetoresistance memory in antiferromagnetic MnTe. *Nature Communications 2016 7:1* **7**, 1–7 (2016).

19. Manchon, A., Li, Q., Xu, L. & Zhang, S. Theory of laser-induced demagnetization at high temperatures. *Phys Rev B Condens Matter Mater Phys* **85**, 064408 (2012).

20. Yin, G. *et al.* Planar Hall Effect in Antiferromagnetic MnTe Thin Films. *Phys Rev Lett* **122**, 106602 (2019).

# Supplementary Notes

**S1. Four-temperature Model**

This model uses temperature to characterize the thermalized distribution function of these particles/quasiparticles, reflecting the energy redistribution among the sub-systems, caused by the multiple types of scattering, e.g. electron-phonon scattering and electron-magnon scattering. We omit the magnon-phonon scattering in the energy transfer since it originates from spin-orbit coupling and has a high-order effect compared with the coupling mechanisms considered[1,2]. The ultrafast regime describes the population of different sub-systems (electrons, magnons and phonons) at non-equilibrium state, where the corresponding temperatures of electrons and magnons is higher than the lattice temperature, thus leading to elevated population of non-equilibrium phonons. The coherent regime where the excitation is in phase with the light or electromagnetic radiation that creating it occurs at the beginning at a faster time scale, differentiate from our focus. We focus on the relaxation process at a slower time scale, e.g. picosecond, where the thermalized distribution function can statistically describe the subsystems of electrons, phonons and magnons[3].

As the pump laser hits the sample, electrons absorb energy from photons and respond immediately to the laser, represented by an energy source term P(t). The transient process is incoherent, starting with the laser excited hot electrons, which in turn emit the hot magnons and phonons at non-equilibrium, through electron-magnon and electron phonon scattering[4]. The electrons interplay directly with the photon, thus manifest an immediate response to the laser, occupying bands of higher energies[1]. These hot electrons statistically reflect an elevated temperature of electron, in the Fermi-Dirac distribution $\frac{1}{\exp\frac{E_{nk}-E_F}{k_B T_e}+1}$, where $E_{nk}$ is the electron energy of band index n at k point, $E_F$ is the Fermi level. The hot electrons consequently excite nonequilibrium magnons, which in turn drive the ultrafast dynamics detected by the probe. We include the spin temperatures of two modes: edge magnon mode and deep magnon mode, to effectively account for the distinct *k*-dependent magnon properties[4]. However, currently there does not exist a reliable theoretical model to provide such *k*-dependent magnon property, and therefore such description is beyond our present capabilities. Thus, we effectively treat the subsystems of the hot magnons through considering their individual thermalized distribution, e.g. two temperatures, and choose the heat capacities of the magnons to reproduce the features of the ultrafast dynamics.

The electron-phonon scattering redistributes energy from the electron to phonon, following the energy and momentum conservation[1]. It is difficult to create a large population of non-equilibrium phonons due to its large thermal conductivity[5], where the elevated temperature is in the order of 10 K, observable through time-domain thermoreflectance[6,7]. The hot phonons as of a boson particle, follow Bose-Einstein distribution $n_{p(m)} = [\exp\left(\frac{\hbar\omega_p}{k_B T_p}\right) - 1]^{-1}$, where $n_p$ is number of phonons, $\hbar\omega_p$ is the intrinsic energy of the phonon (magnon). The elevated temperatures lead to large, wavevector-dependent population of non-equilibrium phonons. Consequently, the hot phonons transduce all the injected energy from the laser to the GaAs substrate at constant temperature $T_{amb}$, which acts as an energy sink.

**S2. QMOKE Probing Mechanism**

The probe laser detects the dynamics of the hot magnons in the temporal domain through Kerr effect. The Kerr angle[8] originates from the difference of the complex refractive indices, $n_\pm$, corresponding to the two eigen modes, $\mathbf{E}_\pm$. In the microscopic framework, the refractive indices, $n_\pm$ are determined by the electric conductivity tensor $\sigma_{\alpha\beta}$,

$$n_\pm^2 = 1 + \frac{4\pi i}{\omega}\left(\frac{\sigma_{yy}+\sigma_{zz}}{2} \pm \sqrt{\frac{1}{4}(\sigma_{yy}-\sigma_{zz})^2 - \sigma_{yz}^2}\right)$$

Where the $\sigma_{yy}$ and $\sigma_{zz}$ are the in-plane directions, and x axis is the out-of-plane direction as of the incident light. The conductivity tensor is approximated using the Kubo formula as

$$\sigma_{\alpha\beta}(\hbar\omega) = \frac{ie^2\hbar}{N_k\Omega_c} \sum_{k,n,m} \frac{f(\epsilon_{mk}) - f(\epsilon_{nk})}{\epsilon_{mk} - \epsilon_{nk}} \frac{\langle\psi_{nk}|v_\alpha|\psi_{mk}\rangle\langle\psi_{mk}|v_\beta|\psi_{nk}\rangle}{\epsilon_{mk} - \epsilon_{nk} - (\hbar\omega + i\eta)}$$

Where $\Omega_c$ is cell volume, $N_k$ is the number of k-points, $f(\epsilon)$ is the Fermi distribution function, and $\eta$ is the smearing parameter. When the wavelength of the probe laser is 800nm, corresponding to the photon energy $\hbar\omega$=1.55 eV, the resonance K points where $\epsilon_{mk} - \epsilon_{nk} = \hbar\omega$ occur at Γ(A) point at the edge of the valence band[8,9]. Vice versa, when the wavelength is 400 nm and the photon energy $\hbar\omega$=3.1 eV, the resonance point is $K_1$ point, deep in the valence band. The magnons coupled to these electrons manifest varying features, e.g. different time characteristics detected by the probe, thus their modes namely are edge mode and deep mode. Phenomenologically, with a lower heat capacity of the band edge magnons compared with that of the deep magnon, the excited magnon number reaches a peak in four picoseconds, where the magnon temperature increases sharply. Consequently, the edge magnon easily decays and their relaxation time is around tens of picoseconds. Vice versa, with a higher heat capacity of the deep band magnons, the emission of magnons is more difficult through the transferred energy from the electrons, which displays a slower increase of temperature. Meanwhile, the relaxation of deeper band magnons takes around hundreds of picoseconds, much longer than the band edge magnons, since the energy flowing from the deeper band magnons to other subsystems has a minor impact on the annihilation of the deeper band magnons. Such wavelength dependence has been observed in static Kerr signal for other magnetic materials[8].

**Reference**

1. Shah, J. Ultrafast Spectroscopy of Semiconductors and Semiconductor Nanostructures. **115**, (1999).
2. Qu, T. & Victora, R. H. Effect of substitutional defects on Kambersky damping in L10 magnetic materials. *Appl Phys Lett* **106**, 72404 (2015).
3. Manchon, A., Li, Q., Xu, L. & Zhang, S. Theory of laser-induced demagnetization at high temperatures. *Phys Rev B Condens Matter Mater Phys* **85**, 064408 (2012).
4. Zhang, W. *et al.* Ultrafast terahertz magnetometry. *Nature Communications 2020 11:1* **11**, 1–9 (2020).

5.  Mu, S. *et al.* Phonons, magnons, and lattice thermal transport in antiferromagnetic semiconductor MnTe. *Phys Rev Mater* **3**, 25403 (2019).

6.  Wilson, R. B. & Cahill, D. G. Anisotropic failure of Fourier theory in time-domain thermoreflectance experiments. *Nat Commun* **5**, (2014).

7.  Chen, J. Y. *et al.* Time-Resolved Magneto-Optical Kerr Effect of Magnetic Thin Films for Ultrafast Thermal Characterization. *Journal of Physical Chemistry Letters* **7**, 2328–2332 (2016).

8.  Higo, T. *et al.* Large magneto-optical Kerr effect and imaging of magnetic octupole domains in an antiferromagnetic metal. *Nat Photonics* **12**, 73 (2018).

9.  Argyres, P. N. Theory of the Faraday and Kerr Effects in Ferromagnetics. *Physical Review* **97**, 334 (1955).